\documentclass[conference]{IEEEtran}
\IEEEoverridecommandlockouts

\usepackage{cite}
\usepackage[dvipdfmx]{graphicx}
\usepackage{lipsum,graphicx,subcaption}
\usepackage{float}
\usepackage{mathtools}  
\usepackage{amsmath}
\usepackage{amssymb}
\usepackage{booktabs}

\usepackage{algorithm}
\usepackage{algorithmic}

\begin{document}
\title{Multi-Stream LDPC Decoder on GPU of Mobile Devices}
\author{
\begin{tabular}{c} Roohollah~Amiri \\ Electrical and Computer Engineering \\ Boise State University, Idaho, USA\\ roohollahamiri@u.boisestate.edu \end{tabular} \and
\begin{tabular}{c} Hani~Mehrpouyan \\ Electrical and Computer Engineering \\ Boise State University, Idaho, USA\\ hanimehrpouyan@boisestate.edu \end{tabular} 
\thanks{978-1-7281-0554-3/19/\$31.00©2019 IEEE}}
\maketitle

\begin{abstract}

Low-density parity check (LDPC) codes have been extensively applied in mobile communication systems due to their excellent error correcting capabilities. However, their broad adoption has been hindered by the high complexity of the LDPC decoder. Although to date, dedicated hardware has been used to implement low latency LDPC decoders, recent advancements in the architecture of mobile processors have made it possible to develop software solutions. In this paper, we propose a multi-stream LDPC decoder designed for a mobile device. The proposed decoder uses graphics processing unit (GPU) of a mobile device to achieve efficient real-time decoding. The proposed solution is implemented on an NVIDIA Tegra board as a system on a chip (SoC), where our results indicate that we can control the load on the central processing units through the multi-stream structure.

\end{abstract}

\section{Introduction}

Low-density parity check (LDPC) codes were originally proposed by Robert Gallager in 1962~\cite{art_gallager} and rediscovered by MacKay and Neal in 1996~\cite{art_macKay}. LDPC codes have been adopted by a wide range of communication standards such as IEEE 802.11n, 10 Gigabit Ethernet (IEEE 802.3an), Long Term Evolution (LTE), and DVB-S2. Chung and Richardson~\cite{art_shannon} showed that a class of LDPC codes could approach the Shannon limit to within 0.0045 dB. However, the error correcting strength of LDPC codes comes at the cost of very high decoding complexity~\cite{art_ldpc_cpu1}. Moreover, to date, there are no closed-form solutions to determine the performance of LDPC codes in various wireless channels and systems. Thus, performance evaluation is typically carried out via simulations on computers or dedicated hardwares~\cite{art_memory_coalesced}.

Since LDPC decoders are computationally-intensive and need powerful computer architectures to result in low latency and high throughput, to date, most LDPC decoders are implemented using application-specific integrated circuits (ASIC) or field-programmable gate array (FPGA) circuits~\cite{art_ldpc_OpenCl_1}. However, their high speed often comes at a price of high development cost and low programming flexibility~\cite{art_convolutional}. Further, it is very challenging to design decoder hardware that supports various standards and multiple data rates~\cite{art_cuda_openmp}. Decoding of LDPC codes is implemented via belief propagation also known as sum-product algorithm (SPA). One advantage of iterative schemes based on the SPA is that it could be parallelized based on the architecture of the code graph~\cite{art_shannon}. In recent years, researchers have used multi-core architectures such as CPUs~\cite{art_cpu_gpu, art_ldpc_cpu0}, graphics processing units (GPUs)~\cite{art_memory_coalesced, art_massively, art_optimize_0}, and advanced RISC machines (ARMs)~\cite{art_ldpc_cpu0, art_neon} to develop high throughput and low latency software-defined radio (SDR) applications. Therefore, designers have recently focused on software implementations of LDPC decoders on multi/many-core devices~\cite{art_massively} to meet the performance requirements of current communication systems.

In microarchitectures, increasing clock frequencies to obtain faster processing performance has reached the limits of silicon-based architectures. Hence, to achieve gains in processing performance, other techniques based on parallel processing is being investigated~\cite{art_ldpc_cpu1}. Todays' multi-core architectures support single instruction multiple data (SIMD), single program multiple data (SPMD), and single instruction multiple threads (SIMT). The general purpose multi-core processors homogeneously replicate a single core, typically with an x86 instruction set, and provide shared memory hardware mechanisms~\cite{art_massively}. Such multi-core structures can be programmed at a high level by using different software technologies~\cite{art_multicore_techs} such as Open Multi-Processing (OpenMP)~\cite{art_openMp_book} which provides a practical and relatively straightforward approach for general-purpose programming.
On the other hand, newer microarchitectures are trying to provide larger SIMD units for vector processing like streaming SIMD extensions (SSE), advanced vector extensions (AVX), and AVX2~\cite{art_intel_sse} on Intel Architectures. In~\cite{art_ldpc_cpu1}, the authors have used Intel SSE/AVX2 SIMD units to implement a high throughput LDPC decoder efficiently. Meanwhile, the power consumption of x86 implementations is incompatible with most of the embedded mobile systems, which makes them useful for simulation purposes only.

Over the last decade, the performance of GPUs has significantly improved mainly due to the demands for visualization technology in the gaming industry. Recent GPUs are composed of many cores which are driven by considerable memory bandwidth. Therefore, they are also being targeted for solving computationally intensive algorithms in a multithreaded and highly parallel fashion. Hence, researchers in the high-performance computing field are applying GPUs to general-purpose applications (GPGPU). Pertaining to the field of communication, researchers have used Compute Unified Device Architecture (CUDA) from NVIDIA \cite{art_gpu_0,art_cuda_openmp, art_memory_coalesced, art_optimize_0, art_layered1} and Open Computing Language (OpenCL)~\cite{art_ldpc_OpenCl} platforms to develop LDPC decoders on GPUs. As an example, the authors in~\cite{art_gpu_0} have achieved almost $1$~Gbps of decoding throughput for LDPC codes on GPU devices. Although these works can achieve extremely high throughputs, their latency beyond seconds, their high power consumption, and their cost make them incompatible with embedded mobile devices. The devices of the end users usually have limited access to a large power source. As such, these devices must operate on limited resources as small processors, tiny memory, and low power budget. In other words, the limited available resources must be used most effectively and efficiently.

ARM-based SDR systems have been proposed in recent years~\cite{art_neon, art_ldpc_cpu0} with the goal of developing an SDR based LDPC decoder that provides high throughput and low latency on a low-power embedded system. The authors in~\cite{art_neon} have used the ARM processor's SIMD and SIMT programming models to implement an LDPC decoder. This approach allows reaching high throughput while maintaining low-latency. However, the proposed ARM-based solution in~\cite{art_neon} is based on the assumption that the ARM processor is solely used for LDPC decoding. However, mobile devices need to support multiple applications simultaneously, and the processing resources cannot be extensively dedicated to the LDPC decoder. Moreover, recent works in SDR LDPC embedded systems are missing the fact that today's mobile devices have powerful CUDA enabled GPUs which can play a significant role as a computing resource in an embedded system. 

This paper proposes a GPU-based LDPC decoder for an embedded device. The structure of the proposed decoder is based on multiple data streams which first makes it scalable to other architectures, and second, the process imposed by the decoding can be controlled by choosing the appropriate number of data streams that are sent to the GPU device. Moreover, since the ARM and GPU of an embedded device are collocated on the same die, the latency issues associated with a GPU implementation is limited.

The remainder of the paper is structured as follows. Section~\ref{sec2} briefly introduces the LDPC error correcting codes and their decoding algorithms. Then the proposed heterogeneous algorithm on embedded mobile targets is described in Section~\ref{sec3}. Section~\ref{sec4} gives experimental results and finally, Section~\ref{conclud} concludes the paper.

\section{LDPC codes and their Decoding Processes}\label{sec2}

\begin{figure}[h]
\begin{centering}
\includegraphics[width=1.0\columnwidth]{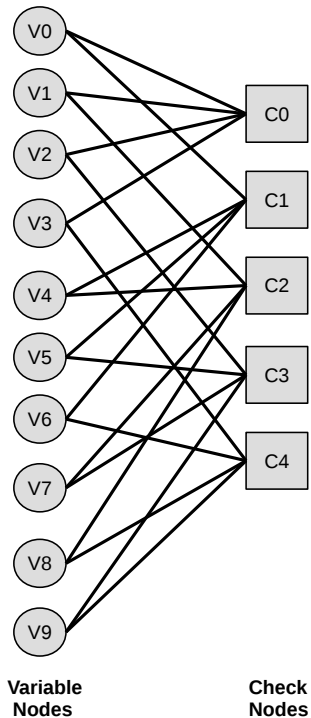}
\caption[width=.3\textwidth]{Tanner graph of the H-matrix in~\eqref{H-matrix}}\label{fig::tanner}
\end{centering}
\end{figure}

LDPC codes are a class of linear block codes with a sparse parity check matrix called H-matrix. Their main advantage is that they provide a performance which is close to that of the channel capacity for various wireless channels. Furthermore, the decoding process of LDPC codes is suited for implementations that make heavy use of parallelism~\cite{art_castello}. Here, we present a brief background on LDPC codes\footnote{The reader is referred to~\cite{book_ldpc} for more information.}. There are two ways to represent LDPC codes. Like all linear block codes, they can be described by their H-matrix, while they can also be represented by a Tanner graph which is a bipartite graph. An LDPC graph consists of a set of variable nodes, a set of check nodes, and a set of edges E. Each edge connects a variable node to a check node. For example, when the $(i,j)$ element of an H-matrix is '1', the $ith$ check node is connected to the $jth$ variable node of the equivalent Tanner graph. Fig.~\ref{fig::tanner} illustrates the equivalent Tanner graph for a 10 variable nodes and 5 check nodes, $(10,5)$, LDPC code with H-matrix in~\eqref{H-matrix}~\cite{art_castello}.

\begin{equation}\label{H-matrix}
H=
  \begin{bmatrix}
    1 & 1 & 1 & 1 & 0 &0 &0 &0 &0 &0 \\
    1 & 0 & 0 & 0 & 1 &1 &1 &0 &0 &0 \\
    0 & 1 & 0 & 0 & 1 &0 &0 &1 &1 &0 \\
    0 & 0 & 1 & 0 & 0 &1 &0 &1 &0 &1 \\
    0 & 0 & 0 & 1 & 0 &0 &1 &0 &1 &1 \\
  \end{bmatrix}
\end{equation}

The general decoding algorithm of LDPC codes is based on the standard two-phase message passing (TPMP) principle described in \cite{art_massively}. This algorithm works in two phases. In the first phase, all the variable nodes send messages to their neighboring parity check nodes, and in the second phase, the parity check nodes send messages to their neighboring variable nodes. One practical variant of message passing algorithms is Min-Sum algorithm which is preferred by designers~\cite{art_neon}. The general steps taken in the Min-Sum algorithm are provided in Algorithm 1. In Algorithm 1, LLR stands for log-likelihood ratio, $\text{CN}_m$ and $\text{VN}_n$ denote the $m$th check node and the $n$th variable node, respectively.

\begin{algorithm}
\caption{Min-Sum algorithm}\label{algorithm1}
\begin{algorithmic}[1]
\STATE \textbf{Loop 1:} Initialization
\FORALL{$t=1 \rightarrow (\text{Max Iterations})$}
\STATE \textbf{Loop 2:} LLR of message $\text{CN}_m$ to $\text{VN}_n$
\STATE \textbf{Loop 3:} LLR of message $\text{VN}_n$ to $\text{CN}_m$
\STATE \textbf{Loop 4:} Hard decision from soft-values
\ENDFOR
\end{algorithmic}
\end{algorithm}

One major drawback of Algorithm 1 is that Loops $2$ and $3$ are updated by separate processing and passed to each other iteratively. It means that the update loop of the variable nodes does not start until all check nodes are updated. This characteristic affects the efficiency of parallel implementation of such an algorithm. 

Due to the poor parallel mapping of the Min-Sum algorithm, more efficient schedules, such as horizontal layered-based decoding algorithm, are proposed which allow updated information to be utilized more quickly in the algorithm, thus, speeding up decoding~\cite{art_layered0, art_layered1}. In fact, the H-matrix can be viewed as a layered graph that is decoded sequentially. The work in~\cite{art_gpu_0} has applied a form of layered belief propagation to irregular LDPC codes to reach $2$ times faster convergence for a given error rate. By using this method, the memory bits usage is reduced by $45\%$ to $50\%$. The layered decoding algorithm is denoted as Algorithm 2 and can be summarized as follows:
\begin{enumerate}
\item All values for the check node computations are computed using variable node messages linked to them.
\item Once, a check node is calculated, the corresponding variable nodes are updated immediately after receiving messages.
\item This process is repeated to the maximum number of iterations.
\end{enumerate}



In this paper, we propose a multi-stream structure for implementing the layered decoding of LDPC codes on the GPU device of a mobile processor with high throughput and low latency performance. By using GPU device as the processing unit, significantly fewer resources of the ARM processor is used for decoding compared to similar work in~\cite{art_neon}. Thus, the ARM processor gains more processing power for other applications running on the device. On the other hand, since the GPU and ARM of a mobile device are sitting on the same die, the latency issues in~\cite{art_gpu_0} are improved.
\section{Algorithm Mapping}\label{sec3}

An efficient implementation of the layered decoding algorithm is a challenging task. The concerning programming drawbacks of this algorithm are as follows:
\begin{enumerate}
\item The number of computations for the number of memory access is low.
\item The data reuse between consecutive computations is low.
\item It requires a large set of random memory access due to the sparse nature of the H-matrix~\cite{art_ldpc_cpu1}.
\end{enumerate}
Therefore, a software-based decoder should take advantage of different parallelism levels offered by the target architecture to achieve high throughput efficiency. In this section, we detail the different parallelism levels, target architecture and the structure of the proposed algorithm.

\subsection{Parallelism Levels in the Proposed Algorithm}
To achieve high throughput performance, a software-based LDPC decoder has to exploit computational parallelism for taking advantage of multi-core architectures. Different parallelism levels are present in a layered decoding algorithm, which include:
\begin{enumerate}
  \item First parallelism level is located inside the check node computations. Executing such computations in parallel is possible. However, this atomic parallelism level is hard to exploit due to the low complexity of computations. On the other hand, two check node computations can be done in parallel if there is no data dependency. Since this is rarely true, this level is hard to exploit and inefficient.
\item Second parallelism level is located at the frame level (complete execution of Algorithm 2). The same computation sequence is executed over consecutive frames. This approach provides an efficient parallel processing algorithm. 
\end{enumerate}
Hence, here, we use the SIMD programming model to decode $F$ frames in parallel. In subsection~\ref{subsec_proposed} the parallel decoding of $F$ frames is referred to as kernel 2 for the sake of simplicity.

\subsection{Data Interleaving/Deinterleaving}
\begin{figure}[H]
\begin{centering}
\includegraphics[width=1.0\columnwidth]{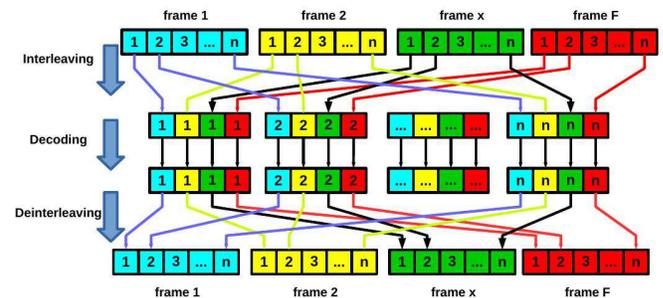}
\caption[width=.3\textwidth]{Data interleaving/deinterleaving process~\cite{art_ldpc_cpu1}}
\label{fig_intreleave}
\end{centering}
\end{figure}

Recall that the implementation of the parallel frame processing is subject to massive irregular memory access due to the structure of H-matrix. To process the same $\text{VN}_n$ element of the $F$ frames at the same time, non-contiguous memory access would affect performance. To solve this issue, a data interleaving process has to be performed before and after the decoding stage to ensure that each set of $F$ frames are reordered to achieve an aligned memory data structure. We use the same procedure as in~\cite{art_ldpc_cpu1} and the reordering is shown in Fig.~\ref{fig_intreleave}. In the proposed structure, interleaving and deinterleaving of frames are called kernel $1$ and kernel $3$.

\subsection{Multi Stream Parallelism}\label{subsec_proposed}

\begin{figure}[h]
\begin{centering}
\includegraphics[width=.80\columnwidth]{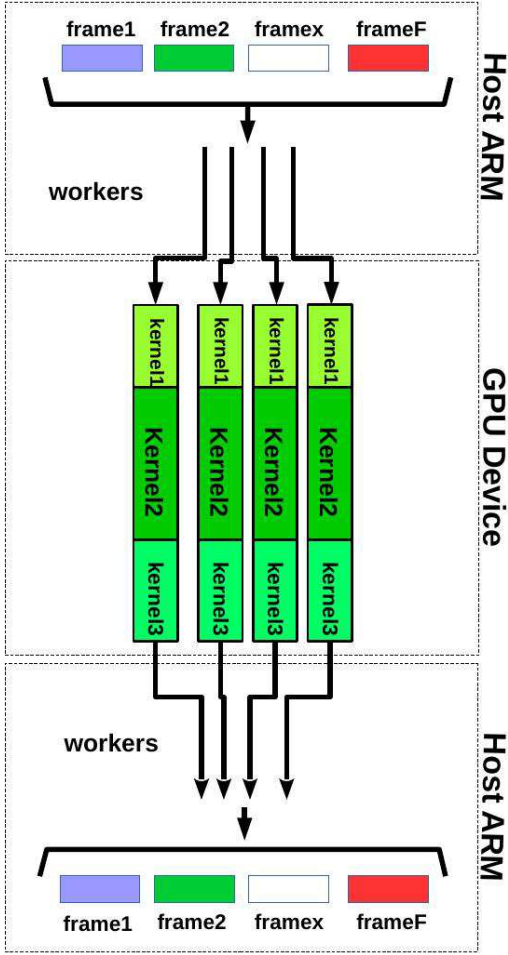}
\caption[width=.3\textwidth]{LDCP decoder data flow}
\label{fig_total}
\end{centering}
\end{figure}

The SIMT programming model is used to decode $W$ sets of $F$ frames concurrently, with $W$ denoting the number of concurrent streams on the GPU device. This multi-core programming is specified by the CUDA API. Each GPU stream is controlled by a \textit{pthread} called \textit{worker} on the host machine (which is an ARM in this case). Each \textit{worker} is responsible for its own sets of frames. By using stream-based processing, the system can decode $W\times F$ frames at the same time. The whole LDPC decoder system model is shown in Fig.~\ref{fig_total}.

\section{Experimental Results} \label{sec4}

The experiments were carried out by decoding LDPC codes using NVIDIA Tegra K1 SoCs and various other structures to show scalability. The programs were compiled via GCC-4.8 and CUDA 6.5. The TK1 is composed of 4 Cortex-A15 ARM processors and one NVIDIA Kepler "GK20a" GPU with 192 SM3.2 CUDA cores. The host platform uses a GNU/Linux kernel 3.10.40.

\begin{figure}[h]
\begin{centering}
\includegraphics[width=0.8\columnwidth]{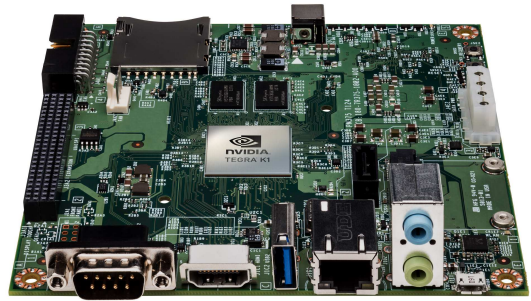}
\caption[width=.3\textwidth]{Tegra-TK1 development board}
\label{tegra}
\end{centering}
\end{figure}

\subsection{Performance Evaluation of the Proposed Algorithm} 
\begin{figure}[H]
\begin{centering}
\includegraphics[width=1.0\columnwidth]{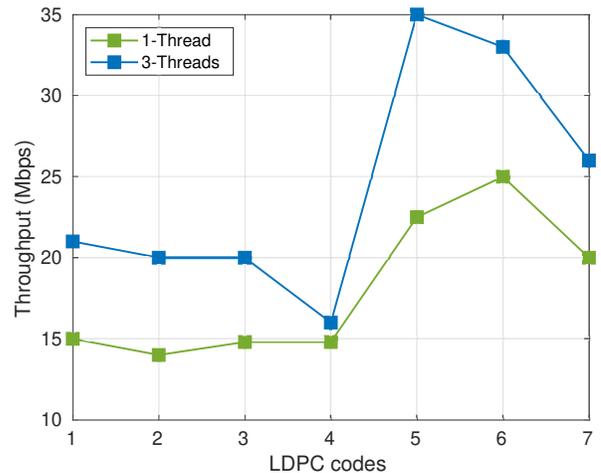}
\caption[width=.5\textwidth]{Measured throughputs for 10 layered decoding iterations ($1-7$ LDPC codes: $576 \times 288, 1024 \times 512, 1200 \times 600, 1944 \times 722, 4000 \times 2000, 8000 \times 4000, 9972 \times 4086$)}\label{fig::air}
\end{centering}
\end{figure}

The first set of experiments evaluates the decoding throughput of different LDPC codes. The codes have different frame lengths: $576$ to $9972$. The results are provided in Fig.~\ref{fig::air} when one or three threads are used to handle one or three GPU streams. Measurements are performed for LDPC decoders that execute $10$ layered-base decoding iterations.

One stream decoding achieves $25$ Mbps, while with three streams it can be as high as $35$ Mbps. For a $\left(4000,2000\right)$ LDPC code and one thread, data transfer takes about $2\times 2.4$ ms, interleaving steps need about $2\times 5$ ms and decoding takes about $150$ ms. For the same code with $3$ threads, data transfer takes approximately $2\times 2.4$ ms, interleaving steps need about $2\times 5$ ms and decoding takes about $150$ ms. Therefore, by introducing more streams to GPU device, its performance does not degrade. In comparison, the latency, i.e., the time for data transfer between the host and GPU device in~\cite{art_gpu_0} is about $20$ ms, is reduced to $4.8$ ms because of the architecture of the embedded mobile device. On the other hand, with introducing three streams to GPU, its processing capacity is used more effectively which results to about $30\%$ throughput improvement in most of our experiments.

\subsection{Performance Comparison with Related Works}
To demonstrate the efficiency of the proposed ARM decoder, its throughput was compared to the ARM related work in~\cite{art_neon}. In~\cite{art_neon}, ARM SIMD units are used to perform vector data processing in parallel frame decoding. In the experiment, the throughput of the proposed decoder is compared to that of~\cite{art_neon} while using $1$ thread for the work in~\cite{art_neon} and $3$ threads in the proposed algorithm. This selection is motivated by the fact that the $1$ thread from~\cite{art_neon} uses a $100\%$ of a core while the $3$ threads for the proposed algorithm only uses $8\%$ of each core resulting in an overall utilization of $24\%$. $10$-iteration decoding performed on Tegra-K1 board gives us the results as shown in Table~\ref{table_compare_arm}. The work in~\cite{art_neon} can achieve much higher throughputs by using more threads on the ARM processor, but by introducing each thread, the whole capacity of one more ARM core is used for decoding. In Table~\ref{table_compare_arm}, it is shown that the proposed algorithm can achieve the similar throughput to that of~\cite{art_neon} when using $24\%$ of ARM processing power and using its GPU device. Although, by using more powerful GPU device, the algorithm can achieve much higher throughputs which has been shown in next subsection. This shows that the proposed algorithm is scalable across platforms.
\noindent
\begin{table}[h]
\centering
\caption{Throughput (Mbps) Comparison With Related Work} \label{table_compare_arm}
\resizebox{\columnwidth}{!}{%
\begin{tabular}{@{}|c|c|c|c|c|@{}}
\toprule
\textbf{}            & \multicolumn{2}{c|}{\textbf{ARM decoder \cite{art_neon}, 1 thread}} & \multicolumn{2}{l|}{\textbf{Proposed decoder, 3 threads}} \\ \midrule
\textbf{code}        & \textbf{(Mbps)}          & \textbf{Processes used}          & \textbf{(Mbps)}               & \textbf{Processes used}              \\ \midrule
\textbf{(4000,2000)} & \textbf{35}              & \textbf{100\%}            & \textbf{34.5}                 & \textbf{\textit{24\%}}                 \\ \midrule
\textbf{(8000,4000)} & \textbf{34}              & \textbf{100\%}            & \textbf{33}                   & \textbf{\textit{24\%}}                \\ \bottomrule
\end{tabular}%
}
\end{table}

\subsection{Performance Comparison on Different GPU Devices}

\begin{figure}
\centering
  \begin{subfigure}[t]{0.5\textwidth}
    \includegraphics[width=1.0\columnwidth]{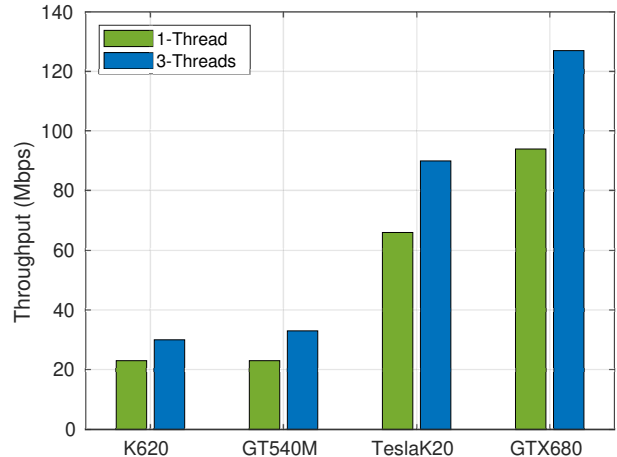}
    \caption{code=(576,288)}
    \label{fig:throu_10_a}
  \end{subfigure}
  \begin{subfigure}[t]{0.5\textwidth}
    \includegraphics[width=1.0\columnwidth]{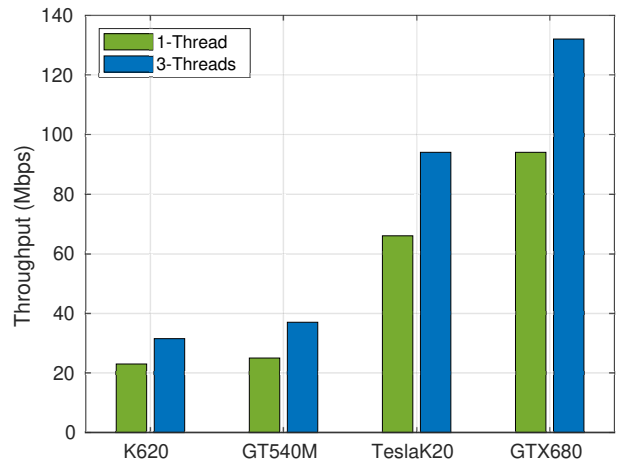}
    \caption{code=(2304,1152)}
    \label{fig:throu_10_b}
  \end{subfigure}
    \begin{subfigure}[t]{0.5\textwidth}
    \includegraphics[width=1.0\columnwidth]{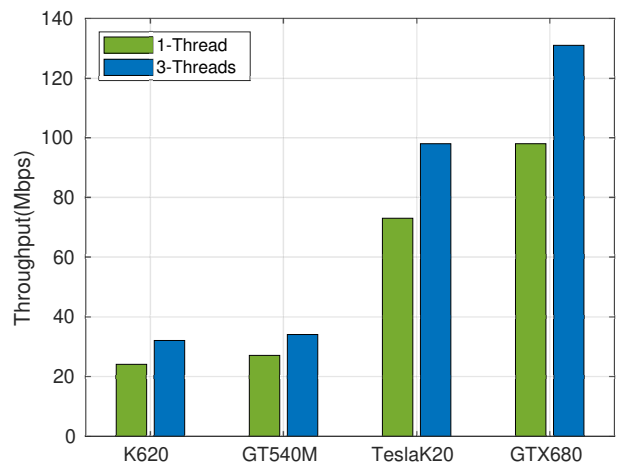}
    \caption{code=(4000,2000)}
    \label{fig:throu_10_c}
  \end{subfigure}
  \caption{10 iteration experiment}\label{fig:10iter}
\end{figure}

\begin{figure}
\centering
  \begin{subfigure}[t]{0.5\textwidth}
    \includegraphics[width=1.0\columnwidth]{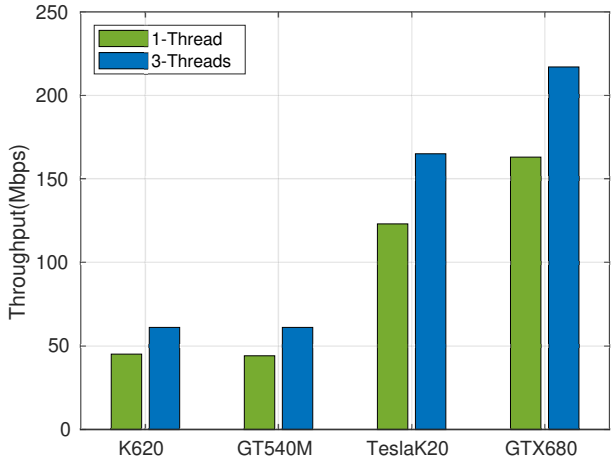}
    \caption{code=(576,288)}
    \label{fig:throu_5_a}
  \end{subfigure}
  \begin{subfigure}[t]{0.5\textwidth}
    \includegraphics[width=1.0\columnwidth]{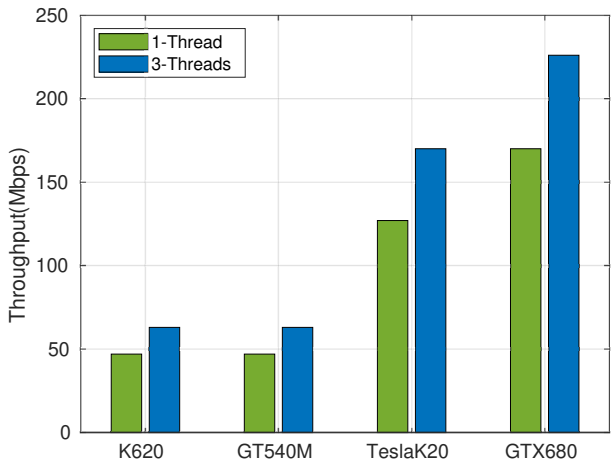}
    \caption{code=(2304,1152)}
    \label{fig:throu_5_b}
  \end{subfigure}
    \begin{subfigure}[t]{0.5\textwidth}
    \includegraphics[width=1.0\columnwidth]{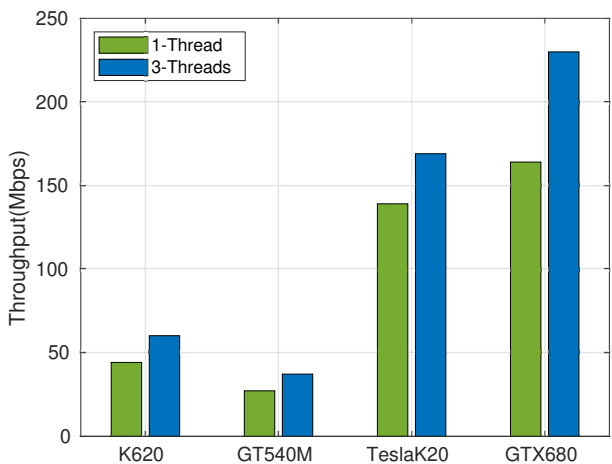}
    \caption{code=(4000,2000)}
    \label{fig:throu_5_c}
  \end{subfigure}
  \caption{5 iteration experiment}\label{fig:5iter}
\end{figure}

GPU devices have different characteristics such as the number of stream multiprocessors, CUDA cores, and working frequencies. A GPU based algorithm should have the scalability to use all the processing capability of a GPU device. The proposed algorithm has been executed on multiple GPU devices. GT540M and K620 are considered as mid-range and GTX680, and TeslaK20 are considered as high power GPU devices. The algorithm is executed for three code lengths as $\left(576,288\right)$, $\left(2304,1152\right)$ and $\left(4000,2000\right)$. The performance is shown for $10$ and $5$ iterations in two sets of figures in Fig.~\ref{fig:10iter} and Fig.~\ref{fig:5iter}. These figures show that the proposed algorithm can achieve up to $230$ Mbps performance across devices. In these set of experiments, an x86 CPU processor is the host.

\section{Conclusion}\label{conclud}
A stream-based approach for GPU-based LDPC decoding on embedded devices was introduced in this paper. This algorithm is based on running multiple concurrent kernels on GPU devices to utilize their processing capacity and freeing up resources on the ARM processor of mobile devices. Our results show that this approach helps to achieve higher throughputs on embedded mobile devices. Experimental results demonstrate that the proposed algorithm is scalable and can achieve high throughputs on multiple GPU devices. Moreover, the proposed algorithm structure provides a trade-off for the operating system to choose between performance and resource management by selecting various values for the number of streams that are used for decoding.
\newpage

\bibliographystyle{IEEEtran}
\bibliography{IEEEabrv,finalProject_bibliography}

\end{document}